\documentclass[sn-mathphys,Numbered]{sn-jnl}

\usepackage{graphicx}%
\usepackage{multirow}%
\usepackage{amsmath,amssymb,amsfonts}%
\usepackage{amsthm}%
\usepackage{mathrsfs}%
\usepackage[title]{appendix}%
\usepackage{xcolor}%
\usepackage{textcomp}%
\usepackage{manyfoot}%
\usepackage{booktabs}%
\usepackage{algorithm}%
\usepackage{algorithmicx}%
\usepackage{algpseudocode}%
\usepackage{listings}%

\usepackage{epsfig}
\usepackage{mathptmx} 
\usepackage{times} 
\usepackage{amsmath} 
\usepackage{amssymb} 
\theoremstyle{remark}
\newtheorem{relation}{Relation}
\usepackage{footmisc}
\usepackage{cancel}
\usepackage{subfigure}

\begin{document}

\title[Correction to the Euler Lagrange Multirotor Model with Euler Angles Generalized Coordinates]{Correction to the Euler Lagrange Multirotor Model with Euler Angles Generalized Coordinates}

\author*[1]{\fnm{Simone} \sur{Martini}}\email{Simone.Martini@du.edu}

\author*[1]{\fnm{Kimon P.} \sur{Valavanis}}\email{Kimon.Valavanis@du.edu}

\author[1]{\fnm{Margareta} \sur{Stefanovic}}\email{Margareta.Stefanovic@du.edu}

\author[2]{\fnm{Matthew J.} \sur{Rutherford}}\email{Matthew.Rutherford@du.edu}

\author[3]{\fnm{Alessandro} \sur{Rizzo}}\email{Alessandro.Rizzo@polito.it}

\affil*[1]{\orgdiv{ECE Department}, \orgname{D. F. Ritchie School of Engineering and Computer Science, University of Denver}, \orgaddress{\city{Denver}, \postcode{80210}, \state{CO}, \country{USA}}}

\affil[2]{\orgdiv{CS Department}, \orgname{D. F. Ritchie School of Engineering and Computer Science, University of Denver}, \orgaddress{\city{Denver}, \postcode{80210}, \state{CO}, \country{USA}}}

\affil[3]{\orgdiv{DET}, \orgname{Politecnico di Torino}, \orgaddress{\city{Torino}, \country{Italy}}}

\abstract{This technical note proves analytically how the exact equivalence of the Newton-Euler and Euler-Lagrange modeling formulations as applied to multirotor UAVs is achieved. This is 
done by deriving a revised Euler-Lagrange multirotor attitude dynamics model. A review of the published literature reveals that the commonly adopted Euler-Lagrange multirotor dynamics model is equivalent to the Newton-Euler model only when it comes to the position dynamics, but not in the attitude dynamics. Step-by-step derivations and calculations are provided to show how modeling equivalence to the Newton-Euler formulation is proven. The modeling equivalence is then verified by obtaining identical results in numerical simulation studies. Simulation results also illustrate that when using the revised model for feedback linearization, controller stability at high gains is improved.}

\keywords{Multirotor, Modeling, Control}

\maketitle

\section{INTRODUCTION}

Derivation of an accurate mathematical model is essential and prerequisite to model-based control of complex dynamic systems in general, and of multirotor UAVs in particular. For example, when considering dynamic inversion control, an accurate mathematical model of the system under consideration should allow for compensation of (any) nonlinear effects, and for linearization of the system dynamics. However, as it happens in almost all cases, it is not realistic to expect complete `capture' of all nonlinear system dynamics effects through a mathematical model, even though the derived model itself does contribute to achieving desirable performance. 

When focusing on multirotor UAVs, the Newton-Euler (N-E) and Euler-Lagrange (E-L) formulations are the two main modeling approaches, albeit following different principles; that is, balancing of forces, and the principle of least action, respectively. Regardless, the N-E and E-L formulations are equivalent, and when applied and implemented on multirotor UAVs, they should return identical results.  

However, after a thorough literature review, this work reveals that when it comes to the published E-L formulations, when substituting the transformation matrix from the body-fixed frame angular velocity to Euler angle derivatives, the N-E and E-L attitude models are not equivalent - this is shown in detail in Section III. Hence, the motivation and main objective of this paper is twofold: First, derive a revised E-L (r-E-L) attitude dynamics model for multirotors, registering at the same time the main differences with the E-L formulations used in literature. The proposed r-E-L model is based on the one presented in \cite{gaull2019rigorous}.  Second, prove analytically the r-E-L model's equivalence to the N-E  formulation (position and attitude dynamics).  Then, to show the implementation improvements with the adoption of the r-E-L, the modeling equivalence is demonstrated through numerical simulations on quadrotors. 

The quadrotor is studied because it is the most widely used configuration of a multirotor UAV. Its mathematical model description may be found in  \cite{luukkonen2011modelling,bouabdallah2007design,lee2017trajectory}, where both formulations are detailed. To be specific, the first E-L formulations may be found in \cite{bouabdallah2004pid,castillo2004real,castillo2004stabilization,raffo2008backstepping,raffo2010integral,das2009backstepping}, while details of the  N-E formulation are presented in  \cite{6289431,l2018introduction}. Both formulations have been widely used and have been implemented for model-based control and navigation. 

State of the art E-L quadrotor models with global validity \cite{4542865,from2012explicit,9196705} are not affected by the findings in this paper due to the coordinate-free nature of their formulations. Nevertheless, the Euler angle variant of the E-L quadrotor model is still adopted in literature work \cite{lavin2023controlling,wang2022quadrotor}, which supports further the correction proposed in this paper. The findings of this work are coherent to the general attitude E-L formulation of \cite{bernstein2023deriving}, which was published at the time of writing this paper. Although following different approaches, both papers arrive, independently, at the same result.

The rest of the paper is organized as follows. Section \ref{2} introduces the required notation, and the N-E and E-L quadrotor dynamics models as found in the related literature.  Section \ref{3} details the proposed r-E-L model and proves its equivalence to the existing N-E model found in the literature. Section \ref{4} includes simulation results. Controller performance comparisons between the proposed and existing models verify and illustrate the equivalence between the two modeling approaches as presented in this paper. Lastly, in Section \ref{5}, conclusions are offered.

\section{Notation and Background Information}\label{2}
\subsection{Notation}
Let $I_n$ denote the $n \times n$ identity matrix. Moreover, given vectors $a,b \in \mathbb{R}^{3}$, denote $S(a) b = a\times b$, where 
\begin{eqnarray}
    S(a) = \left [
    \begin{array}{ccc}
0 & -a_3 & a_2\\
a_3 & 0 & -a_1\\
-a_2 & a_1 & 0
    \end{array}
    \right]
\end{eqnarray} 
is the $3\times 3$ skew symmetric matrix composed of the elements of $a$. In addition, consider the unit vector $\mathbf{e_3}= [0,0,1]^T$. The notation $\frac{da}{dt}$ and $\dot{a}$ is used interchangeably throughout the paper. 

In what follows, for clarity purposes, the N-E and E-L quadrotor dynamics are considered, which may be easily generalized to any multirotor UAV dynamics.

\subsection{Quadrotor Nonlinear Dynamics}
The N-E quadrotor dynamics, as presented in \cite{luukkonen2011modelling}, are described by the following equations
\begin{eqnarray} \label{attdyn}
J\dot \omega &=& M - S(\omega)J\omega \label{attitude} \;,\\
\dot v &=& \frac{1}{m}T\mathbf{e_3} - S(\omega)v - gR^T(\eta)\mathbf{e_3}  \label{position}\;,
\end{eqnarray}
where $S(\omega) \in \mathbb{R} ^{3\times 3}$ is the skew symmetric matrix of the angular velocities, $\omega$, defined in the body-fixed frame, $J \in \mathbb{R} ^{3\times 3}$ is the constant diagonal inertia matrix, $M$ is the external torque induced by the quadrotor propellers in the body-fixed frame, $v$ is the quadrotor linear velocity vector in the body-fixed reference frame, $m$ is the total mass of the quadrotor, $g$ is the gravitational acceleration, and $T$ is the total produced thrust. The matrix $R(\eta)\in \mathbb{R}^{3\times 3}$ represents the rotation from the body-fixed frame to the inertial frame. Note that the choice of $R(\eta)$ is not unique since the quadrotor dynamics are invariant to any choice of Euler angles configuration that may be used to represent the attitude of the quadrotor. 

The E-L formulation, as introduced in \cite{Martini}, is represented by the following two equations 
\begin{eqnarray}
\ddot \eta &=& J_R^{-1}(\eta) (M - C(\eta,\dot\eta)\dot \eta)\label{attitude_lag} \;,\\
\ddot p &=& \frac{1}{m}TR(\eta)\mathbf{e_3} - g\mathbf{e_3}\label{pos_lag} \;,
\end{eqnarray}
where $\eta = [\phi,\theta,\psi]^T$ is the vector of any choice of Euler angles configuration, and $p = [x,y,z]^T$ is the inertial reference frame position vector. $J_{R}(\eta)$ is the rotated inertia matrix and $C(\eta,\dot\eta)$ is the matrix accounting for centrifugal and Coriolis effects.\\

\section{Equivalence of the N-E and E-L Modeling Formulations}\label{3}
To verify if the E-L and N-E models are equivalent, a coordinate transformation should lead from one formulation to the other. Starting from the position dynamics, the linear velocity expressed in the body-fixed frame is related to the inertial frame velocity by the following equation
\begin{eqnarray}
    v = R^T(\eta) \dot p \label{rot}
\end{eqnarray}
Thus, by substituting (\ref{rot}) in (\ref{position}), the following equation is derived (with all steps shown in detail)  
\setlength{\arraycolsep}{0.0em}
\begin{eqnarray}\label{eqn:position2}
\frac{d(R^T(\eta) \dot p)}{dt} &{}={}& - S(\omega)R^T(\eta) \dot p - gR^T(\eta)\mathbf{e_3} + \frac{1}{m}T\mathbf{e_3}\nonumber\\
\dot{R^T(\eta)} \dot p + R^T(\eta) \ddot p &{}={}& - S(\omega)R^T(\eta) \dot p - gR^T(\eta)\mathbf{e_3} + \frac{1}{m}T\mathbf{e_3}\nonumber\\
\cancel{S(\omega)^T R^T(\eta) \dot p} + R^T(\eta) \ddot p &{}={}& \cancel{- S(\omega)R^T(\eta) \dot p} - gR^T(\eta)\mathbf{e_3} + \frac{1}{m}T\mathbf{e_3}\nonumber\\
\ddot p &{}={}& - g\mathbf{e_3} + \frac{1}{m}TR(\eta)\mathbf{e_3}
\end{eqnarray}
\setlength{\arraycolsep}{5pt}
The resulting equation is the same as (\ref{pos_lag}), thus, the equivalence between the N-E and E-L position dynamics formulations (as found in the literature) is proven.
\begin{figure}
    \centering
    \includegraphics[width=\columnwidth]{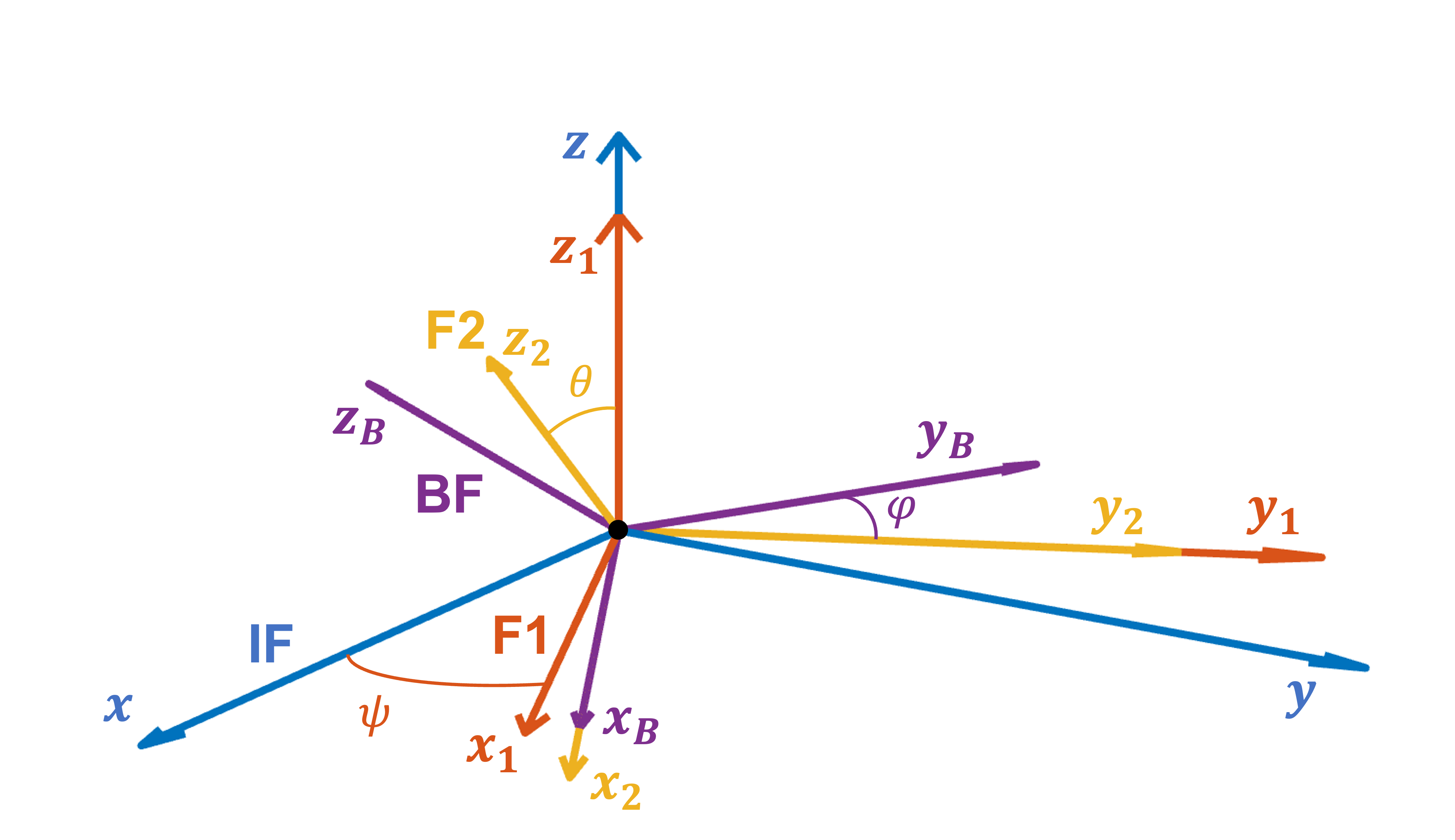}
    \caption{Tait-Bryan $321$ Sequence of Elementary Rotations from \textbf{IF} to \textbf{BF}}
    \label{fig:rot}
\end{figure}

However, the relationship between the Euler angles derivatives $\dot \eta$ and the angular velocities $\omega$ is less intuitive. This relationship may be derived by considering the rotation from the inertial reference frame (\textbf{IF}) to the body-fixed frame (\textbf{BF}) as the composition of three elementary rotations. To achieve this, two intermediate reference frames are defined, \textbf{F1} and \textbf{F2}, respectively. Following \cite{novara}, without loss of generality, to show how this relationship is derived, the Tait-Bryan $321$ sequence of rotations is employed as shown in Fig. \ref{fig:rot}, and it is detailed next.
\begin{itemize}
    \item \textbf{IF}$\longrightarrow$\textbf{F1}\\ \textbf{F1} is obtained from the elementary rotation of an angle $\psi$ with respect to the $z$-axis of the \textbf{IF}, defined as $R_3(\psi)^T = R_3(-\psi)$. The related angular velocity in the \textbf{IF} is $\omega_{\psi,I} = \left[0, 0, \dot\psi \right]^T$. It follows that the angular velocity in \textbf{F1} is given by $\omega_{\psi,1} = R_3(-\psi)\omega_{\psi,I} = \omega_{\psi,I}$.
    \item \textbf{F1}$\longrightarrow$\textbf{F2}\\ \textbf{F2} is obtained from the elementary rotation of an angle $\theta$ with respect to the $y_1$-axis of \textbf{F1}, defined as $R_2(\theta)^T = R_2(-\theta)$. The related angular velocity in \textbf{F1} is $\omega_{\theta,1} = \left[0, \dot\theta, 0 \right]^T$. It follows that the angular velocity in \textbf{F2} is given by $\omega_{\theta,2} = R_2(-\theta)\omega_{\theta,1} = \omega_{\theta,1}$. Moreover, in \textbf{F2}, the angular velocity related to the angle $\psi$ is $\omega_{\psi,2} = R_2(-\theta)\omega_{\psi,1} = R_2(-\theta)\omega_{\psi,I}$.
    \item \textbf{F2}$\longrightarrow$\textbf{BF}\\ \textbf{BF} is obtained from the elementary rotation of an angle $\phi$ with respect to the $x_2$-axis of \textbf{F2}, defined as $R_1(\phi)^T = R_1(-\phi)$. The related angular velocity in \textbf{F2} is $\omega_{\phi,2} = \left[\dot\phi, 0, 0 \right]^T$. It follows that the angular velocity in \textbf{BF} is given by $\omega_{\phi,B} = R_1(-\phi)\omega_{\phi,2} = \omega_{\phi,2}$. Moreover, in \textbf{BF}, the angular velocity related to the angle $\theta$ is $\omega_{\theta,B} = R_1(-\phi)\omega_{\theta,2} = R_1(-\phi)\omega_{\theta,1}$. Finally, in \textbf{BF}, the angular velocity related to the angle $\psi$ is $\omega_{\psi,B} = R_1(-\phi)\omega_{\psi,2} = R_1(-\phi)R_2(-\theta)\omega_{\psi,I}$.
\end{itemize}

Therefore, the angular velocity as expressed in the \textbf{BF} is
\begin{eqnarray}\label{eqn:W}
\omega \!\!\!&\!=\!&\!\!\! \omega_{\phi,B} + \omega_{\theta,B} + \omega_{\psi,B}\nonumber = \omega_{\phi,2} + R_1(-\phi)\omega_{\theta,1} + R_1(-\phi)R_2(-\theta)\omega_{\psi,I} \;\;\;= \nonumber\\
\!\!\!&\!=\!&\!\!\! \left[\!\!\!\!\begin{array}{c}\dot\phi\\0\\0\end{array}\!\!\!\!\right]\! + R_1(-\phi)\left[\!\!\!\!\begin{array}{c}0\\\dot\theta\\0\end{array}\!\!\!\!\right] + R_1(-\phi)R_2(-\theta)\left[\!\!\!\!\begin{array}{c}0\\0\\\dot\psi\end{array}\!\!\!\!\right] = I_{3_1} \dot\phi + A_2 \dot\theta + B_3 \dot\psi = \underbrace{\left[I_{3_1}, A_2, B_3 \right]}_{=W(\eta)} \left[\!\!\!\!\begin{array}{c}\dot\phi\\\dot\theta\\\dot\psi\end{array}\!\!\!\!\right]\nonumber\\
\end{eqnarray}
where $I_{3_1}$ is the first column of the $3\times3$ identity matrix, $A_2$ is the second column of $A = R_1(-\phi)$, and $B_3$ is the third column of $B = R_1(-\phi)R_2(-\theta)$.\\
Thus, in more compact form  
\begin{eqnarray}
\label{InvKin}
    \omega = W(\eta) \dot \eta \\
    \dot \eta = W^{-1}(\eta) \omega \label{InvKin2}
\end{eqnarray}
where $W$ depends on the choice of the Euler angle sequence. One example, for clarification purposes, may be found in \cite{beard2008quadrotor}. Kinematic relations (\ref{InvKin}) and (\ref{InvKin2}) are invariant to the choice of the Euler angles, but the configuration needs to be consistent with the choice of $R$. However, different Euler angles configurations will result in a different $W$, and so, given (\ref{InvKin2}), it is recommended to choose a configuration in which $W$ is invertible for the whole flight envelope, see \cite{Martini}, except when the pitch angle is equal to $\pi$ (an uncommon state outside acrobatic manoeuvres).

When substituting (\ref{InvKin}) in the N-E attitude model (\ref{attitude}), this leads to 
\begin{eqnarray}
\label{eqn: attitude2}
\begin{aligned}
M &= \frac{d(J{\left(W(\eta) {\dot\eta} \right)})}{dt} + S(W(\eta) \dot \eta) JW(\eta) \dot \eta \nonumber\\
M &= J \dot W(\eta) \dot{\eta} + JW(\eta) \ddot{\eta}  + S(W(\eta) \dot \eta) JW(\eta) \dot \eta \nonumber\\
M &= JW(\eta) \ddot{\eta} + (J \dot W(\eta) + S(W(\eta) \dot \eta) JW(\eta)) \dot \eta
\end{aligned} 
\end{eqnarray}
where $S(W \dot \eta)$ is the skew symmetric matrix of $W(\eta) \dot \eta$.\ 

When compared to the literature E-L model shown in (\ref{attitude_lag}), which can be rewritten as
\begin{eqnarray}\label{el_model_liter}
M = J_R(\eta) \ddot \eta + C(\eta,\dot\eta)\dot \eta
\end{eqnarray}
it leads to the following inequalities
\begin{eqnarray}
    &J_R(\eta)\ddot\eta = W^T(\eta) JW(\eta) \ddot\eta\neq JW(\eta)\ddot\eta\label{ineq1}\\
    &C(\eta,\dot \eta) \dot\eta= \left(\dot {J_R} - \frac{1}{2} \frac{\partial({\dot \eta}^T J_R)}{\partial \eta}\right)\dot\eta \neq \left( J \dot W(\eta) + S(W(\eta) \dot \eta)JW(\eta)\right)\dot\eta \label{ineq2}
\end{eqnarray}

and, in general, contrary to what should be expected, the multirotor attitude dynamics following the E-L and N-E formulations found in the literature do not produce an equivalent result given that 
\begin{eqnarray}
J\dot \omega - S(\omega)J\omega \neq J_R(\eta) \ddot \eta + C(\eta,\dot\eta)\dot \eta
\end{eqnarray}

However, this shortcoming may be rectified based on the approach introduced in \cite{gaull2019rigorous}, in which a proof is provided for the equivalence of the projective N–E equations and the E-L equations of second kind for spatial rigid multibody systems. 

Considering the proof in  \cite{gaull2019rigorous}, it is shown that for multirotor UAVs (including quadrotors), the attitude dynamics E-L equations should be written in the following form  
\begin{eqnarray}
\label{lag_eq_true}
    \frac{d}{dt} \frac{\partial L}{\partial \dot \eta} - \frac{\partial L}{\partial \eta} = W^T(\eta) M
\end{eqnarray}
instead of the one presented in \cite{luukkonen2011modelling}, that is 
\begin{eqnarray}\label{lag_eq_false}
    \frac{d}{dt} \frac{\partial L}{\partial \dot \eta} - \frac{\partial L}{\partial \eta} = M
\end{eqnarray}
with the Lagrangian, $L = \frac{1}{2} \omega ^T J \omega$, which, in the case of attitude dynamics, is equivalent to the rotational kinetic energy.

An intuitive reason for writing the r-E-L form as (\ref{lag_eq_true}) is given by noticing that a premultiplication of $W^T$ would result in (\ref{ineq1}) and (\ref{ineq2}) being equalities. Moreover, this approach is similar to the quaternion variant of the E-L formulation as shown in \cite{6564793}. Building on the above observation, it is now shown that the same steps presented in \cite{gaull2019rigorous} may be followed to model multirotor UAVs / quadrotors. 

Since it is considered that the forces are applied to the center of mass of the quadrotor (multirotor), the position and attitude dynamics can be analyzed independently. Hence, only the E-L equations for the angular dynamics are derived. 

To prove the equivalence of the r-E-L model, to the N-E model, the following relations are defined first. For better readability, $W$ is written without explicit dependence on $\eta$.
\vspace{\baselineskip}\begin{relation}
Consider $W(\eta)^{-1}$, the rows of which are given by $w_{inv,1}, w_{inv,2}, w_{inv,3}$, as follows
\begin{eqnarray}
    W^{-1} = \left(\begin{array}{c}w_{inv,1}\\ w_{inv,2}\\ w_{inv,3}\end{array} \right)
\end{eqnarray}
Next, define the matrix
\setlength{\arraycolsep}{0.0em}
\begin{eqnarray}\label{skewW}
\Sigma(W^{-1})\!=\!\left(\!\begin{array}{c}
         \frac{\partial w_{inv,1}^{T}}{\partial \eta}W^{-1} \\
         \frac{\partial w_{inv,2}^{T}}{\partial \eta}W^{-1} \\
         \frac{\partial w_{inv,3}^{T}}{\partial \eta}W^{-1} \\
    \end{array}\! \right)\!-\! \left(\!\begin{array}{c}
         (\frac{\partial w_{inv,1}^{T}}{\partial \eta}W^{-1})^T \\
         (\frac{\partial w_{inv,2}^{T}}{\partial \eta}W^{-1})^T \\
         (\frac{\partial w_{inv,3}^{T}}{\partial \eta}W^{-1})^T \\
    \end{array}\! \right) =\left(\!\begin{array}{c}
         S(w_{inv,1}) \\
         S(w_{inv,2}) \\
         S(w_{inv,3}) \\
    \end{array}\! \right)
\end{eqnarray}
\setlength{\arraycolsep}{5pt}

This matrix, for any Euler angle sequence, is composed of the skew symmetric matrices of $S(w_{inv,i})$ with $i = 1,2,3$.
\end{relation}
\vspace{\baselineskip}\begin{relation} Note that the time derivative of $W^{-1}$ can be expressed as
\setlength{\arraycolsep}{0.0em}
\begin{eqnarray}\label{Wdot}
    \frac{dW^{-1}}{dt}  {=} \!\left(\!\begin{array}{c}
         \frac{\partial w_{inv,1}^{T}}{\partial \eta}\dot \eta \\
         \frac{\partial w_{inv,2}^{T}}{\partial \eta}\dot \eta \\
         \frac{\partial w_{inv,3}^{T}}{\partial \eta}\dot \eta \\
    \end{array}\! \right) \! {=} \! \left(\!\begin{array}{c}
         \dot \eta^T(\frac{\partial w_{inv,1}^{T}}{\partial \eta})^T \\
         \dot \eta^T(\frac{\partial w_{inv,2}^{T}}{\partial \eta})^T \\
         \dot \eta^T(\frac{\partial w_{inv,3}^{T}}{\partial \eta})^T \\
    \end{array}\! \right) \! {=} \! \left(\!\begin{array}{c}
         \omega^T(\frac{\partial w_{inv,1}^{T}}{\partial \eta}W^{-1})^T \\
         \omega^T(\frac{\partial w_{inv,2}^{T}}{\partial \eta}W^{-1})^T \\
         \omega^T(\frac{\partial w_{inv,3}^{T}}{\partial \eta}W^{-1})^T \\
    \end{array}\! \right)
\end{eqnarray}
\setlength{\arraycolsep}{5pt}
\end{relation}
\vspace{\baselineskip}\begin{relation}\label{rel3} The following is obvious
\setlength{\arraycolsep}{0.0em}
\begin{eqnarray}
    \frac{\partial \eta}{\partial \eta} = \frac{\partial \dot \eta}{\partial \dot \eta} = \frac{\partial (W^{-1}\omega)}{\partial \dot \eta} = W^{-1}\frac{\partial \omega}{\partial \dot \eta} = I
\end{eqnarray}
\setlength{\arraycolsep}{5pt}
\end{relation}

\vspace{\baselineskip}\begin{relation}
From Rel. \ref{rel3}, the following holds 
\setlength{\arraycolsep}{0.0em}
\begin{eqnarray}\label{partW}
    \frac{\partial W^{-1}}{\partial \eta} \omega  {=} \left(\begin{array}{c}
          \omega^T(\frac{\partial w_{inv,1}^{T}}{\partial \eta}) \\
          \omega^T(\frac{\partial w_{inv,2}^{T}}{\partial \eta}) \\
          \omega^T(\frac{\partial w_{inv,3}^{T}}{\partial \eta}) \\
    \end{array} \right) {=}  \left(\begin{array}{c}
          \omega^T(\frac{\partial w_{inv,1}^{T}}{\partial \eta})W^{-1} \\
          \omega^T(\frac{\partial w_{inv,2}^{T}}{\partial \eta})W^{-1} \\
          \omega^T(\frac{\partial w_{inv,3}^{T}}{\partial \eta})W^{-1} \\
    \end{array} \right) \frac{\partial \omega}{\partial \dot \eta}
\end{eqnarray}
\setlength{\arraycolsep}{5pt}
\end{relation}
\vspace{\baselineskip}\begin{relation}\label{rel5}
In addition, using Rel. \ref{rel3}, it is true that 
\setlength{\arraycolsep}{0.0em}
\begin{eqnarray}
    \left( \frac{\partial W^{-1} \omega}{\partial \eta}\right) =  \frac{\partial \dot \eta}{\partial \eta} = \frac{d}{dt}\left(\frac{\partial \eta}{\partial \eta}\right) = \frac{d}{dt}\left(W^{-1} \frac{\partial \omega}{\partial \dot \eta}\right)
\end{eqnarray}
\setlength{\arraycolsep}{5pt}
that leads to the following 
\setlength{\arraycolsep}{0.0em}
\begin{eqnarray}
    \left( \frac{\partial W^{-1} \omega}{\partial \eta}\right) = \frac{d}{dt}\left(W^{-1} \right) \frac{\partial \omega}{\partial \dot \eta} + W^{-1}  \frac{d}{dt}\left(\frac{\partial \omega}{\partial \dot \eta}\right)
\end{eqnarray}
\setlength{\arraycolsep}{5pt}
\end{relation}
\vspace{\baselineskip}\begin{relation}\label{rel6}
Rearranging Rel. \ref{rel5} and substituting (\ref{Wdot}), (\ref{partW}), and (\ref{skewW}), the following equality is obtained
\setlength{\arraycolsep}{0.0em}
\begin{eqnarray}
        \frac{d}{dt}\left(\frac{\partial \omega}{\partial \dot \eta}\right)
    && =  W \left( \frac{\partial W^{-1} \omega}{\partial \eta} - \frac{d}{dt}\left(W^{-1} \right) \frac{\partial \omega}{\partial \dot \eta} \right)  = W \left( W^{-1}\frac{\partial \omega}{\partial  \eta} + \frac{\partial W^{-1} }{\partial \dot \eta}\omega - \frac{d}{dt}\left(W^{-1} \right) \frac{\partial \omega}{\partial \dot \eta} \right)= \nonumber\\
    && \: {=}\frac{\partial \omega}{\partial \eta} + W \left[ \!\left(\!\begin{array}{c}
          \omega^T(\frac{\partial w_{inv,1}^{T}}{\partial \eta})W^{-1} \\
          \omega^T(\frac{\partial w_{inv,2}^{T}}{\partial \eta})W^{-1} \\
          \omega^T(\frac{\partial w_{inv,3}^{T}}{\partial \eta})W^{-1} \\
    \end{array} \!\right)\! \!-\!\! \left(\!\begin{array}{c}
         \omega^T(\frac{\partial w_{inv,1}^{T}}{\partial \eta}W^{-1})^T \\
         \omega^T(\frac{\partial w_{inv,2}^{T}}{\partial \eta}W^{-1})^T \\
         \omega^T(\frac{\partial w_{inv,3}^{T}}{\partial \eta}W^{-1})^T \\
    \end{array}\! \right)\! \right] \!\frac{\partial \omega}{\partial \dot \eta} =  \nonumber\\
    && \: {=} \frac{\partial \omega}{\partial \eta} + W \left(\begin{array}{c}
         \omega^T S(w_{inv,1}) \\
         \omega^T S(w_{inv,2}) \\
         \omega^T S(w_{inv,3}) \\
    \end{array} \right) \frac{\partial \omega}{\partial \dot \eta} = \frac{\partial \omega}{\partial \eta} + W \left(\begin{array}{c}
         - w_{inv,1} S(\omega) \\
         - w_{inv,2} S(\omega) \\
         - w_{inv,3} S(\omega) \\
    \end{array} \right) \frac{\partial \omega}{\partial \dot \eta} \nonumber\\ 
    && \: {=} \frac{\partial \omega}{\partial \eta} - \underbrace{WW^{-1}}_{=I_3} S(\omega) \frac{\partial \omega}{\partial \dot \eta} 
\end{eqnarray}
\setlength{\arraycolsep}{5pt}
leading to 
\setlength{\arraycolsep}{0.0em}
\begin{eqnarray}\label{prop1}
    \frac{d}{dt}\left(\frac{\partial \omega}{\partial \dot \eta}\right) = \frac{\partial \omega}{\partial \eta} - S(\omega) \frac{\partial \omega}{\partial \dot \eta}
\end{eqnarray}
\setlength{\arraycolsep}{5pt}
\end{relation}
\vspace{\baselineskip}\begin{relation}\label{rel7}
Considering (\ref{InvKin}) it is easy to show that 
\setlength{\arraycolsep}{0.0em}
\begin{eqnarray}\label{prop2}
    \left( \frac{\partial \omega}{\partial \dot \eta}\right)^T =  \frac{(\partial \dot \eta^TW^T)}{\partial \dot \eta} = W^T
\end{eqnarray}
\setlength{\arraycolsep}{5pt}
\end{relation}\vspace{\baselineskip}\vspace{\baselineskip}
Given the relationships above, it is shown next that the r-E-L formulation leads to an equivalent result with the N-E attitude model.
\vspace{\baselineskip}\vspace{\baselineskip}

\begin{proof}
The multirotor r-E-L formulation (\ref{lag_eq_true}) can be rewritten as 
\begin{eqnarray}
    \frac{d}{dt} ({\frac{\partial \frac{1}{2} \omega ^T J \omega}{\partial \dot \eta}}) - \frac{\partial \frac{1}{2} \omega ^T J \omega}{\partial \eta} = W^T M
\end{eqnarray}
Since $J$ is a constant symmetric matrix, this is equivalent to
\begin{eqnarray}\label{lag_start}
    &\frac{d}{dt} \left[\left( \frac{\partial \omega}{\partial \dot \eta}\right)^T J \omega\right] - {\left(\frac{\partial \omega}{\partial \eta}\right)}^T J \omega = W^T M \nonumber \\
        &\frac{d}{dt} \left[\left( \frac{\partial \omega}{\partial \dot \eta}\right)^T \right] J \omega + \left( \frac{\partial \omega}{\partial \dot \eta}\right)^T J \dot \omega - {\left(\frac{\partial \omega}{\partial \eta}\right)}^T J \omega = W^T M 
\end{eqnarray}
From (\ref{lag_start}), by using Rel. \ref{rel6} and Rel. \ref{rel7} one obtains 
\setlength{\arraycolsep}{0.0em}
\begin{eqnarray}\label{lag_cont}
        &\begin{aligned}
            \left[\cancel{\left(\frac{\partial \omega}{\partial \eta}\right)^T} + \left(\frac{\partial \omega}{\partial \dot \eta}\right)^TS(\omega) \right]& J \omega + \left( \frac{\partial \omega}{\partial \dot \eta}\right)^T J \dot \omega - \cancel{{\left(\frac{\partial \omega}{\partial \eta}\right)}^T J \omega} = W^T M \nonumber
        \end{aligned}\\
       &\quad W^T S(\omega)J \omega + W^T J \dot \omega = W^T M 
\end{eqnarray}
\setlength{\arraycolsep}{5pt}
And, for $W$ having full rank, (\ref{lag_cont}) is rewritten as
\setlength{\arraycolsep}{0.0em}
\begin{eqnarray}
         J \dot \omega + S(\omega)J \omega = M 
\end{eqnarray}
\setlength{\arraycolsep}{5pt}
This corresponds to the N-E quadrotor model formulation. This equivalence does not hold when using the E-L formulation from the literature (\ref{lag_eq_false}). This concludes the proof. 
\end{proof}

Therefore, the proposed attitude r-E-L model is 
\begin{eqnarray}
W^TM = J_R \ddot \eta + C\dot \eta
\end{eqnarray}
which rectifies the literature E-L model (\ref{el_model_liter}) and this result is reconfirms and agrees with findings in \cite{bernstein2023deriving} for the general attitude dynamics.
\section{Comparison with Previous Formulations}\label{4}
Given the proof of the mathematical equivalence of the r-E-L and N-E, it is straightforward to state that the presented formulation leads to improved performance with respect to the E-L quadrotor model found in literature. This section shows the difference in performance considering the model employed in section \ref{comp1} and \ref{comp2}, as simulation platform, and in section \ref{comp3}, for model-based control.
\subsection{Implementation Comparison between the different models}\label{comp1}
In this section, it is shown that the error between the r-E-L and N-E formulations is lower compared to the one between the E-L and N-E formulations. To do so, the same rotor input velocity, $u = [475.9+ 0.1\sin{t}, 476.2+ 0.1\sin{t}, 476, 476.1]$ is applied to the (N-E, E-L, r-E-L) quadrotor models for a period of $60 s$, and the root mean square error (RMSE) is computed for the generalized coordinates ($p$, $\eta$, $\dot p$, $\dot \eta$). The choice of these rotor inputs is such that the quadrotor will cover a long range in $x,y,z$, while having non constant attitude and not reaching any singular configuration. From Table \ref{tab1}, which illustrates results with an integration step of $10 ms$, it is evident that the r-E-L has a RMSE of $3$ to $6$ orders of magnitude smaller than the literature E-L. 

\begin{table}[htb]
\caption{Comparison with E-L Model from Literature (10ms)}\label{tab1}
\begin{tabular}{|c|c|c|}
\hline
      & E-L        & r-E-L \\ \hline
$RMSE_{p}$                  & $1.853$ & $67.976 \times 10^{-6}$  \\ \hline
$RMSE_{\eta}$                  & $8.135 \times 10^{-3}$ & $496.720 \times 10^{-9}$ \\ \hline
$RMSE_{\dot p}$ & $160.833 \times 10^{-3}$ & $6.210 \times 10^{-6}$  \\ \hline
$RMSE_{\dot \eta}$ & $4.416 \times 10^{-3}$ & $21.692 \times 10^{-9}$  \\ \hline
\end{tabular}
\end{table}
Moreover, Table \ref{tab2} shows that by further decreasing the integration step to $1 ms$, the error of the revised model decreases even more, while this is not true for the E-L model found in literature E-L, leading to an error of $6$ to $9$ order of magnitude smaller.
\begin{table}[htb]
\caption{Comparison with E-L Model from Literature (1ms)}\label{tab2}
\begin{tabular}{|c|c|c|}
\hline
      & E-L        & r-E-L \\ \hline
$RMSE_{p}$                  & $1.853$ & $68.297 \times 10^{-9}$  \\ \hline
$RMSE_{\eta}$                  & $8.135 \times 10^{-3}$ & $499.466 \times 10^{-12}$ \\ \hline
$RMSE_{\dot p}$ & $160.828 \times 10^{-3}$ & $6.240 \times 10^{-9}$  \\ \hline
$RMSE_{\dot \eta}$ & $4.436 \times 10^{-3}$ & $21.821 \times 10^{-12}$  \\ \hline
\end{tabular}
\end{table}

Note that, even though they are studied independently, the position dynamics are affected by the attitude dynamics due to the influence of the rotation matrix. For this reason it is worth considering the RMSE for $p$ and $\dot p$. as well.

\subsection{Comparison with Multibody Dynamic Simulator}\label{comp2}
Now, the comparison of these mathematical models with respect to a multibody dynamic simulator such as Mathworks's Simacape Multibody is considered. Note that Simscape Multibody uses its own dynamic engine to solve the equation of motions, while only the structure and the inertia of the quadrotor is provided by the user. 
As shown in Table \ref{tab3}, providing the same input as before to all models, the N-E and the r-E-L state dynamics are almost identical to the one of a multibody dynamic simulator, while the literature E-L model has a much larger error. Again, as shown in Table \ref{tab4}, this becomes even more evident when decreasing the simulation step to $1 ms$.
\begin{table}[htb]
\caption{Comparison with Multibody Dynamic Simulator (10ms)}\label{tab3}
\begin{tabular}{|c|c|c|c|}
\hline
      & N-E & E-L        & r-E-L \\ \hline
$RMSE_{p}$                  & $130.785 \times 10^{-6}$ & $1.853$ & $149.865 \times 10^{-6}$  \\ \hline
$RMSE_{\eta}$                  & $255.423 \times 10^{-6}$ & $8.338 \times 10^{-3}$ & $255.868 \times 10^{-6}$ \\ \hline
$RMSE_{\dot p}$ & $6.535 \times 10^{-6}$ & $160.833 \times 10^{-3}$ & $9.841 \times 10^{-6}$  \\ \hline
$RMSE_{\dot \eta}$ & $461.558 \times 10^{-6}$ & $4.354 \times 10^{-3}$ & $461.559 \times 10^{-6}$  \\ \hline
\end{tabular}
\end{table}

\begin{table}[htb]
\caption{Comparison with Multibody Dynamic Simulator (1ms)}\label{tab4}
\begin{tabular}{|c|c|c|c|}
\hline
      & N-E & E-L        & r-E-L \\ \hline
$RMSE_{p}$                  & $130.965 \times 10^{-6}$ & $1.853$ & $130.968 \times 10^{-6}$  \\ \hline
$RMSE_{\eta}$                  & $25.550 \times 10^{-6}$ & $8.155 \times 10^{-3}$ & $25.551 \times 10^{-6}$ \\ \hline
$RMSE_{\dot p}$ & $6.550 \times 10^{-6}$ & $160.828 \times 10^{-3}$ & $6.551 \times 10^{-6}$  \\ \hline
$RMSE_{\dot \eta}$ & $46.303 \times 10^{-6}$ & $4.429 \times 10^{-3}$ & $46.303 \times 10^{-6}$  \\ \hline
\end{tabular}
\end{table}
It is essential to state that in order to achieve these error magnitudes, all mathematical models (N-E, E-L, r-E-L) need to account for the gyroscopic effect, which is automatically computed in the dynamic simulator. The gyroscopic effect is modeled as in \cite{Martini}, and since it depends on the external input $u$, for the r-E-L model, it is pre multiplied by $W^T$.

\subsection{Effect on Model Based Control}\label{comp3}
\begin{figure}[!htb]
\centering
\includegraphics[width=0.50\columnwidth]{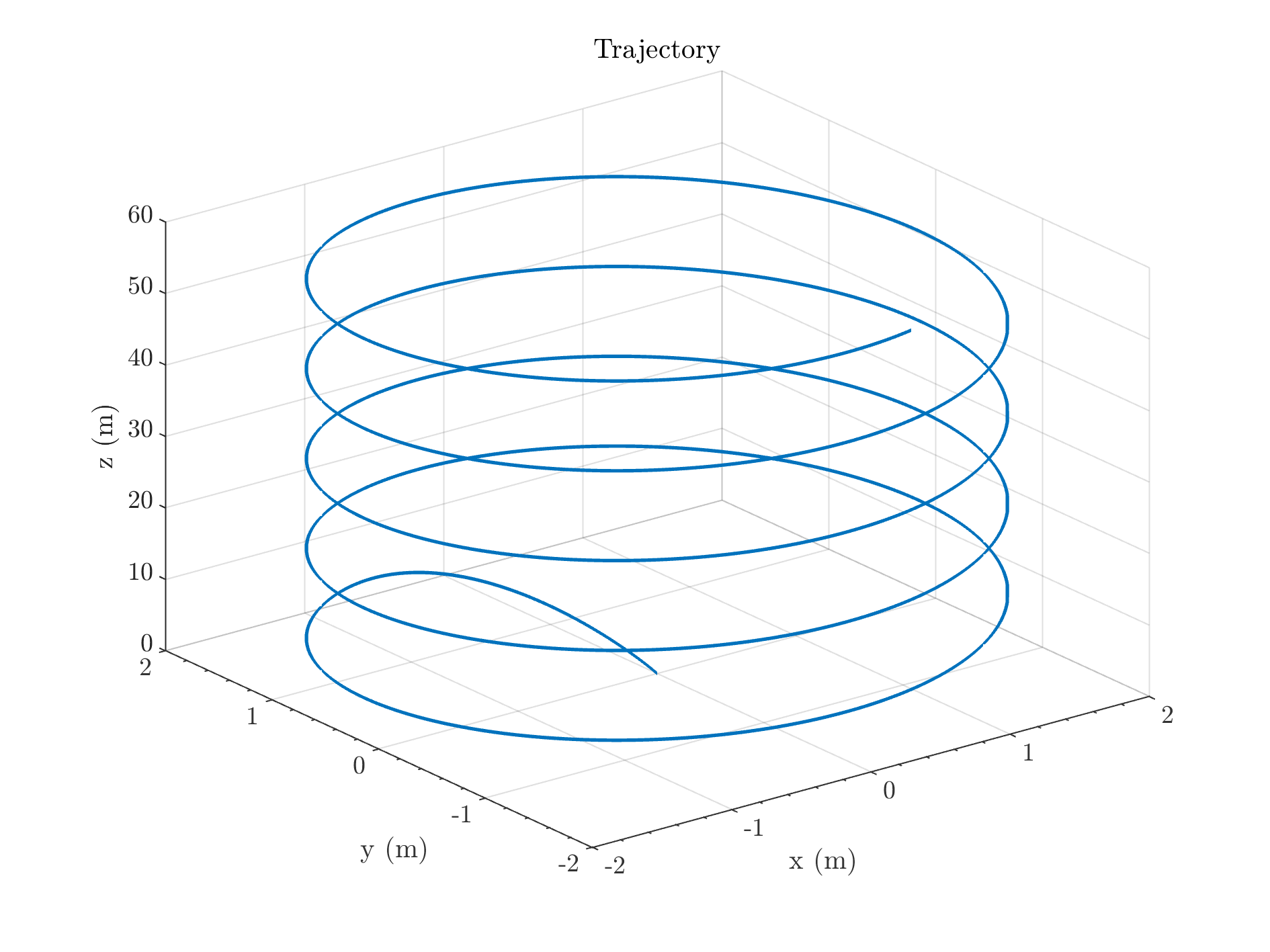}\hfill
\includegraphics[width=0.42\columnwidth]{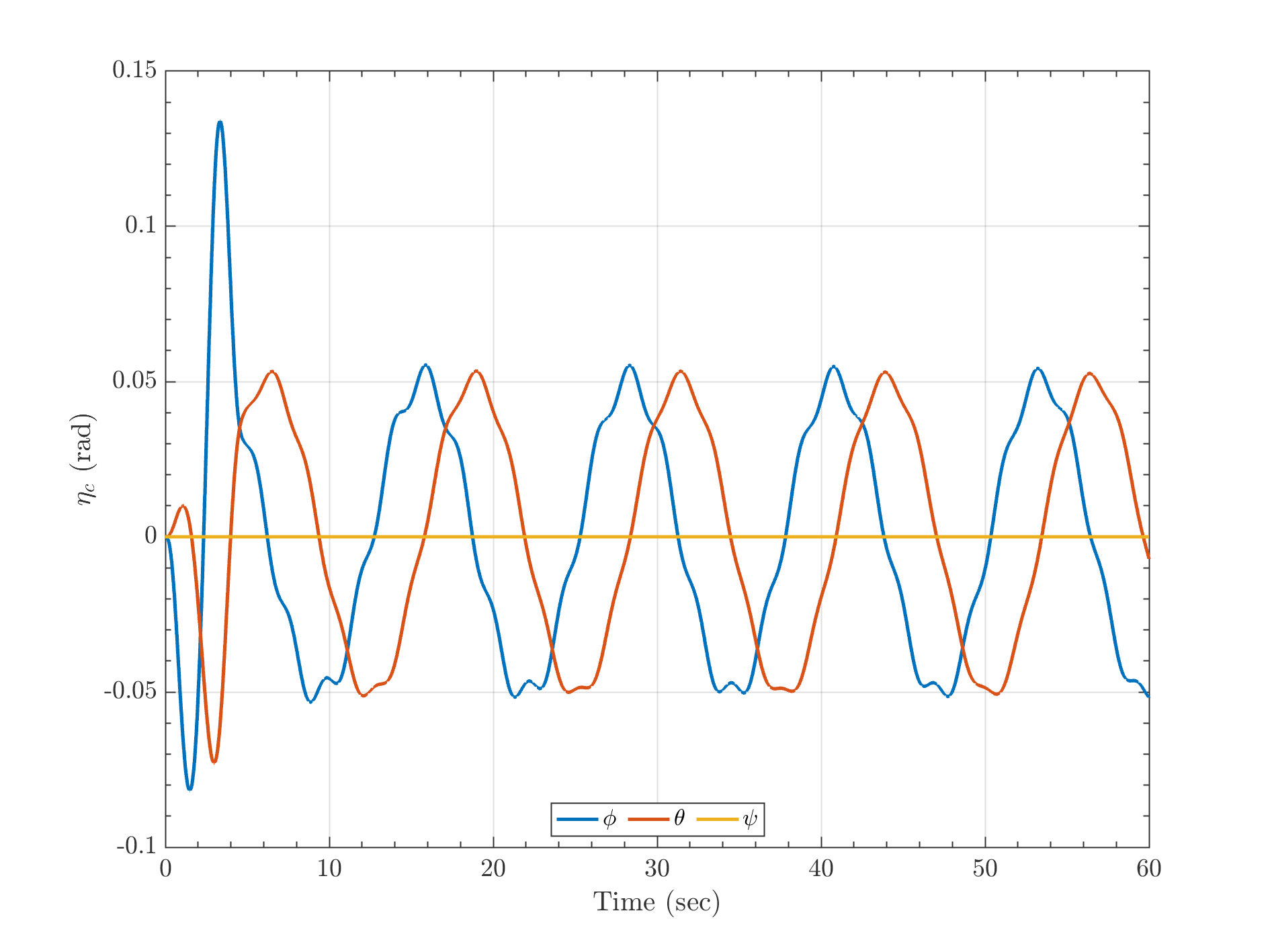}
\caption{\label{fig1}3D Trajectory (left) \& Attitude Trajectory (right)}
\end{figure}

By implementing a PID controller with feedback linearization on the multibody dynamic simulator quadrotor the different effects of the E-L and r-E-L are analyzed.
The helix trajectory to track is displayed in Fig. \ref{fig1}, along the resulting in the attitude trajectory.

\begin{figure}[!htb]
\centerline{\includegraphics[width=0.76\columnwidth]{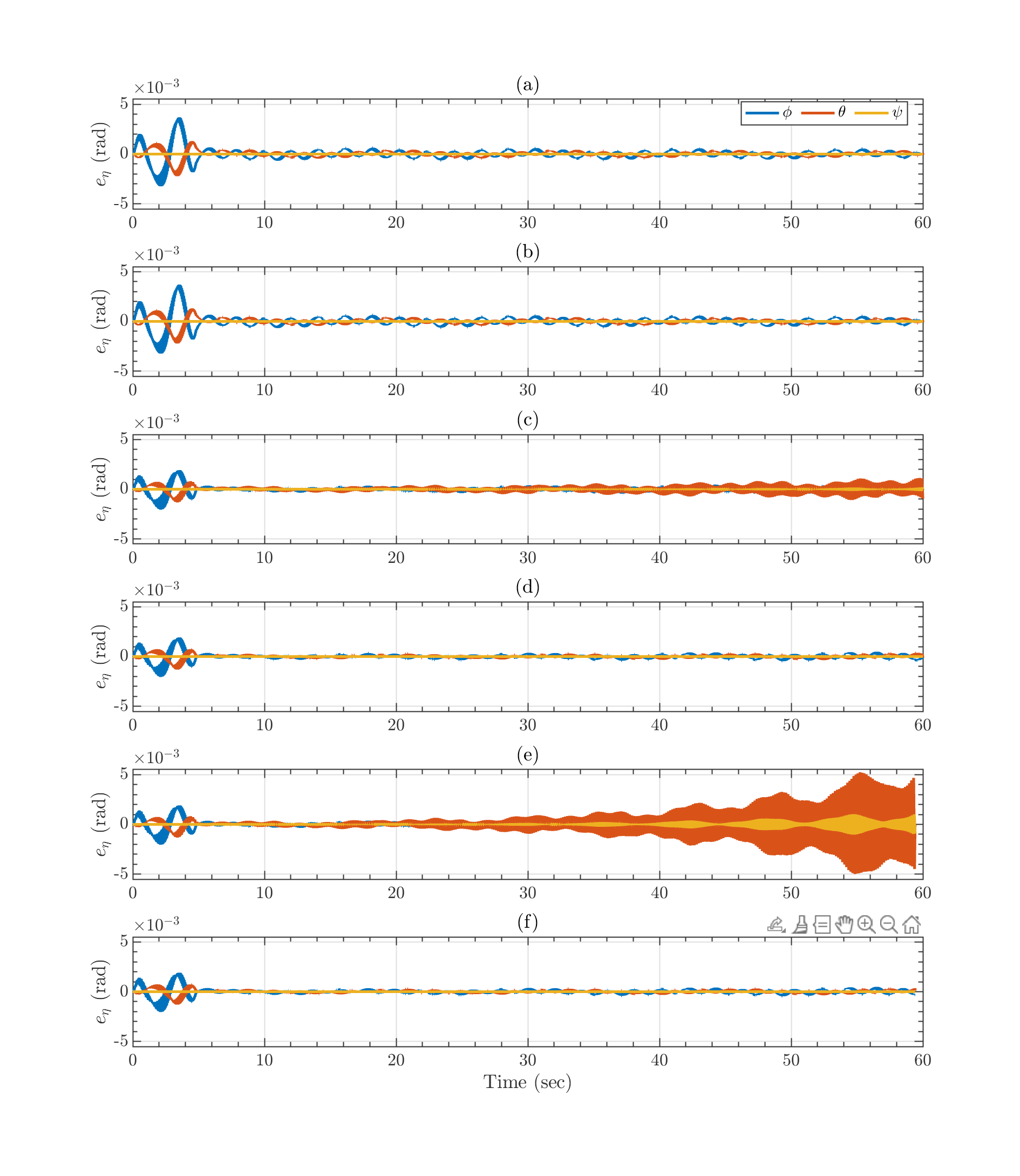}}
\caption{Closed Loop System Attitude Error $e_{\eta}$. (a),(c),(e) use literature E-L for dynamic compensation, (b),(d),(f) use r-E-L for dynamic compensation. Controller gain $Ki = 8\times 10^3,\,15.5\times 10^3,\,16\times 10^3$ respectively}
\label{fig3}
\end{figure}

The attitude errors resulting from the numerical simulations with different PID gain configurations are shown in Fig. \ref{fig3}. While at some gain values the controller performance is identical, it is shown that by increasing the controller gains, the controller that uses the r-E-L for dynamic compensation has bigger margin of stability. The feedback linearization using the literature E-L model leads to unstable behaviour with smaller gain compared to the r-E-L.

\section{Conclusions}\label{5}
A revised E-L attitude dynamics model of quadrotors / multirotor UAVs has been presented and derived. The equivalence to the N-E formulation has been demonstrated through analytical steps and numerical simulations. Compared to existing E-L formulations found in literature, the r-L-E model is equivalent to the N-E model and therefore, when exploited for dynamic compensation, it provides better stability of the closed loop controlled system. Building on this work, previously presented feedback linearization controllers using E-L formulation (including the ones from the authors) could be improved.

\bmhead{Acknowledgments}
This research is conducted at the University of Denver Unmanned Systems Research Institute ($\text{DU}^{2}\text{SRI}$) in collaboration with Politecnico di Torino.

\bibliography{Artical_arxiv}

\end{document}